\documentclass[aps,pra,twocolumn,superscriptaddress,nofootinbib,10pt]{revtex4-2}

\usepackage[T1]{fontenc}
\usepackage{amsmath,amssymb,bm,mathtools}
\usepackage{hyperref}
\usepackage{xcolor}
\usepackage{microtype}
\usepackage{url}
\usepackage{booktabs}
\usepackage{orcidlink}

\newcommand{\R}{\mathbb{R}}
\newcommand{\Rcal}{\mathcal{R}}
\newcommand{\epsr}{\varepsilon_r}
\newcommand{\mur}{\mu_r}
\newcommand{\ii}{\mathrm{i}}
\definecolor{linknavy}{RGB}{25,55,95}
\definecolor{linkred}{RGB}{150,40,55}
\hypersetup{
    colorlinks         = true,
    linkcolor          = blue,
    citecolor          = red,
    urlcolor           = linknavy,
    filecolor          = linknavy,
    menucolor          = linknavy,
    pdfstartview       = FitH,
    pdfpagemode        = UseOutlines,
    bookmarks          = true,
    bookmarksnumbered  = true,
    bookmarksopen      = true,
    bookmarksopenlevel = 2,
    pdfdisplaydoctitle = true,
    pdftitle            = {Exact transformation-optics representation of the Ellis wormhole and limits of finite-device scattering emulation},
    pdfauthor           = {Abdullah Guvendi, Claudio Furtado, and Omar Mustafa},
    breaklinks         = true
}
\allowdisplaybreaks[1]
\newcommand{\dd}{\mathrm{d}}

\begin{document}

\title{Exact transformation-optics representation of the Ellis wormhole and limits of finite-device scattering emulation}

\author{Abdullah Guvendi\orcidlink{0000-0003-0564-9899}}
\email{abdullah.guvendi@erzurum.edu.tr}
\affiliation{Department of Basic Sciences, Erzurum Technical University, 25050 Erzurum, T\"urkiye}

\author{Claudio Furtado\orcidlink{0000-0002-3455-4285}}
\email{furtado@fisica.ufpb.br (corresponding author)}
\affiliation{Departamento de Física, Universidade Federal da Paraíba, Caixa Postal 5008, João Pessoa, Paraíba, 58051-970, Brazil}

\author{Omar Mustafa\orcidlink{0000-0001-6664-3859}}
\email{omar.mustafa@emu.edu.tr}
\affiliation{Department of Physics, Eastern Mediterranean University, 99628, G. Magusa, north Cyprus, Mersin 10 - Türkiye}

\begin{abstract}
We derive a global isotropic representation of the ultrastatic Ellis wormhole and establish which features of its electromagnetic response can, and cannot, be reproduced by a finite device. An orientation-preserving coordinate map sends the signed proper-radial line to a punctured Euclidean chart with refractive index $n(\rho)=1+a^2/(4\rho^2)$. Through the Pleba\'nski correspondence, this gives the exact impedance-matched continuum medium $\boldsymbol{\varepsilon}_r=\boldsymbol{\mu}_r=n(\rho)\mathbf I$ and its full vector-Maxwell operator. For the ideal two-ended geometry, we derive the electromagnetic radial operator, establish its spectral properties, and express its scattering amplitudes through Jost solutions and Wronskians. Nonzero transmission into the second asymptotic end implies $|\mathcal R_{\mathrm{id},L}|<1$ at every real frequency, whereas a regular, lossless finite one-port has unit-modulus reflection. The reverse triangle inequality therefore yields a strictly positive lower bound on the complex reflection-amplitude mismatch. This theorem applies to the stated lossless single-port class and is not extended to lossy or multiport realizations. We compute converged multipole scattering data and establish the low-frequency threshold scaling $|\mathcal T_L|^2\propto(\omega a/c)^{4L+2}$, with coefficients independently validated for $L=1,2,3,4$. The transmission cross section tends to $\pi a^2$, while the ideal profile's total scattering cross section has a logarithmic forward divergence. We also construct a smooth finite-shell target and derive an exact finite-band constitutive-error formula for a calibrated passive Lorentz model. The constitutive benchmark is kept distinct from scattering equivalence throughout.
\end{abstract}


\maketitle
\section{Introduction}
\label{sec:introduction}

Transformation optics (TO) exploits the coordinate covariance of Maxwell's equations to turn a prescribed geometry for wave propagation into the constitutive parameters of a material. The idea goes back to the coordinate formulation of refraction by Ward and Pendry~\cite{WardPendry1996}, and it took its modern form with Leonhardt's conformal mapping~\cite{Leonhardt2006} and the field-control approach of Pendry, Schurig, and Smith~\cite{Pendry2006}. The material parameters and ray paths in transformation media were worked out soon afterwards~\cite{Schurig2006}, full-wave simulations of cloaks followed~\cite{Cummer2006}, and a metamaterial cloak was demonstrated at microwave frequencies~\cite{SchurigScience2006}. The field has since been reviewed from several angles: applications beyond ray control~\cite{ChenChanSheng2015}, theory and design~\cite{ThompsonCummer2012}, conformal constructions~\cite{XuChen2015}, open problems~\cite{McCallRoadmap2018}, and the practical limits of cloaking~\cite{McCall2013Cloaking}. Broader accounts of the link between TO and the geometry of light are also available~\cite{LeonhardtPhilbin2009,ChenChanSheng2010,LeonhardtPhilbin2010}. One point recurs throughout this literature and matters here: a coordinate transformation that is exact in the continuum need not have a bounded, causal realization of finite size.

Reading a metric as a medium has a long history. Gordon introduced an effective optical metric for light in moving media~\cite{Gordon1923}, and Pleba\'nski showed that Maxwell's equations on a gravitational background are equivalent to flat-space equations in an effective medium~\cite{Plebanski1960}. This optical view of gravity was developed further by de Felice~\cite{Felice1971} and by Leonhardt and Philbin~\cite{LeonhardtPhilbin2006}. Practical designs, however, need not reproduce a spacetime metric exactly. Devices built from ideal thin lenses form a separate design class~\cite{Courtial2018RayOpticalTO}, which has led to ideal-lens cloaks~\cite{Belin2019IdealLensCloaks} and to shifty cloaks that behave like optical wormholes~\cite{Courtial2024ShiftyCloaks}. Conformal mapping has also been used to emulate transfer between Riemann sheets as an electromagnetic-wormhole analogue~\cite{Sun2017RiemannSheets}. These ray-optical and analogue constructions are useful precedents, but none of them is a global, three-dimensional, full-wave representation of the Ellis metric.

The Ellis wormhole is an ultrastatic, spherically symmetric geometry with a signed proper-radial coordinate and two asymptotically flat ends~\cite{Ellis1973}. Traversable wormholes and their energy conditions are discussed by Morris and Thorne~\cite{MorrisThorne1988}, in Visser's monograph~\cite{Visser1995}, and in the collection edited by Lobo~\cite{Lobo2017}. Waves in wormhole backgrounds are not a new subject. Cl\'ement analyzed Klein--Gordon and Maxwell scattering by the Ellis geometry~\cite{Clement1984}; P\'erez Bergliaffa and Hibberd reduced electromagnetic propagation through a static wormhole to a one-dimensional scattering problem~\cite{PerezBergliaffaHibberd2000}; Konoplya and Zhidenko computed quasinormal modes and grey-body factors of scalar and electromagnetic fields for traversable wormholes of arbitrary shape, a family that contains the Ellis shape function, using higher-order WKB methods~\cite{KonoplyaZhidenko2010}; and wave optics and lensing by Ellis wormholes have been studied with the scalar wave equation~\cite{NambuNodaSakai2019,YooHaradaTsukamoto2013}. A related two-dimensional problem, the curvature-induced potential on a catenoid, has also been examined~\cite{Dandoloff2010Catenoid}. More recently, a general parametrization of wormhole spacetimes has been applied to their shadows and quasinormal modes~\cite{BronnikovKonoplyaPappas2021}, and transmission through wormholes is surveyed in a recent review of grey-body factors~\cite{KonoplyaZhidenko2026Review}. We therefore do not claim to introduce electromagnetic scattering by an Ellis wormhole, nor the computation of its grey-body factors. Our aim is different: to build its global constitutive representation and to ask which of its scattering properties a finite device can reproduce. The central question is a wave-physics one: which exact scattering features of a two-ended continuum geometry are compatible with the channel structure of a finite optical realization? By combining the global Maxwell representation with a quantitative scattering analysis and a finite-device mismatch theorem, we identify a precise boundary between continuum equivalence and realizable one-port response.

Related ray- and wave-optics analyses of optical-wormhole geometries include an elliptic optical wormhole and a minimal-surface wormhole~\cite{GuvendiOpticalWormhole2025,GuvendiMinimalSurface2025}. These studies provide the closest context for the present work while addressing geometries and optical constructions distinct from the global four-dimensional Ellis metric and source-free Maxwell system considered here. Complementary studies have developed observational tests of static and spherically symmetric wormholes through the general-relativistic Poynting--Robertson effect and ray tracing~\cite{DeFalco2021PRD}, as well as reconstruction methods for wormhole solutions in curvature-based extended gravity~\cite{DeFalco2021EPJC}.

We construct and verify a global, orientation-preserving isotropic map of the signed Ellis radial line and derive the associated impedance-matched constitutive tensors together with the full vector-Maxwell operator. On this basis, we formulate the electromagnetic scattering problem across the two asymptotic ends, establish the spectrum of the radial operator, and compute converged multipole amplitudes and cross sections, including the low-frequency threshold law. We then prove a strictly positive reflection-mismatch bound for regular, lossless finite one-port realizations, and introduce a smooth finite-shell regularization for which the worst-case constitutive error of a calibrated passive Lorentz response is obtained exactly. Taken together, these results separate the exact continuum representation and ideal-geometry scattering from the material approximation achievable in a finite device. The analysis is situated within work on transformation-optics wormholes~\cite{Greenleaf2007}, singular transformations in inverse problems~\cite{GreenleafLassasUhlmann2003}, and their mathematical and physical limits~\cite{GreenleafReview2009}. Near-cloaking performance is a boundary-value problem~\cite{BaoLiu2014,BaoLiuZou2014}; passive cloaks are subject to bandwidth bounds~\cite{MonticoneAlu2016}, while gain-assisted loss compensation constitutes a distinct implementation issue~\cite{Han2011GainAssistedTO}. Accordingly, exact identities, numerical results, and device-level implications are identified separately throughout the paper.

The paper is organized as follows. Section~\ref{sec:geometry} derives the global isotropic map and verifies its range and metric transformation. Section~\ref{sec:maxwell} obtains the constitutive tensors and full vector-Maxwell operator from the Pleba\'nski correspondence. Section~\ref{sec:puncture} identifies the puncture as the second asymptotic end at infinite optical distance. Section~\ref{sec:scattering} derives the electromagnetic radial equation, establishes its spectral properties, and presents the converged scattering data and cross sections. Section~\ref{sec:finite} proves the lossless one-port mismatch bound and constructs a finite-shell Lorentz benchmark, keeping constitutive and scattering errors distinct. Section~\ref{sec:conclusions} summarizes the results and their precise scope.

\section{Ellis geometry and the global isotropic map}
\label{sec:geometry}

We adopt signature $(-+++)$, SI coordinate time $t$, and explicit $c$. The ultrastatic Ellis line element is given in the original drainhole construction by Ellis~\cite{Ellis1973}:
\begin{align}
\dd s^2&=-c^2\dd t^2+\dd l^2+(l^2+a^2)\dd\Omega^2,
\label{eq:2:1}\\
\dd\Omega^2&=\dd\theta^2+\sin^2\theta\,\dd\phi^2,
\qquad a>0,\quad l\in\R .
\label{eq:2:2}
\end{align}
Here $l$ is signed proper-radial distance. The areal radius is
\begin{equation}
\Rcal(l)=\sqrt{l^2+a^2},\quad
\Rcal'=\frac{l}{\Rcal},\quad
\Rcal''=\frac{a^2}{\Rcal^3}.
\label{eq:2:3}
\end{equation}
Thus $\Rcal$ has a unique nondegenerate minimum $\Rcal(0)=a$, and the throat area is $A_{\rm th}=4\pi a^2$. The limits $l\to\pm\infty$ are distinct asymptotically flat ends, both of which are retained by the signed coordinate.

In the coordinate ordering $(t,l,\theta,\phi)$, the metric determinant is
\begin{equation}
g_l=-c^2(l^2+a^2)^2\sin^2\theta,
\qquad
\sqrt{-g_l}=c(l^2+a^2)\sin\theta .
\label{eq:2:4}
\end{equation}
The factor $c$ in the volume density follows from $g_{tt}=-c^2$; using $x^0=ct$ instead sets $g_{00}=-1$ and removes this coordinate factor without changing the spacetime volume form.

We seek a global coordinate $\rho>0$ for which the spatial metric is conformally Euclidean:
\begin{equation}
\dd l^2+(l^2+a^2)\dd\Omega^2
=n^2(\rho)\left(\dd\rho^2+\rho^2\dd\Omega^2\right).
\label{eq:2:5}
\end{equation}
Independent matching of radial and angular coefficients gives
\begin{equation}
\left(\frac{\dd l}{\dd\rho}\right)^2=n^2(\rho),
\qquad
l^2+a^2=n^2(\rho)\rho^2.
\label{eq:2:6}
\end{equation}
Choosing the orientation-preserving branch $\dd l/\dd\rho=n>0$ and eliminating $n$ yields
\begin{equation}
\frac{\dd\rho}{\rho}=\frac{\dd l}{\sqrt{l^2+a^2}}.
\label{eq:2:7}
\end{equation}
Integration, with a dimensionless logarithm and a constant fixed by the positive-end normalization $\rho/l\to1$ as $l\to+\infty$, gives
\begin{equation}
\operatorname{arsinh}(l/a)=\ln(2\rho/a).
\label{eq:2:8}
\end{equation}
The exact map and inverse are therefore
\begin{equation}
\rho=\frac{l+\sqrt{l^2+a^2}}{2},
\qquad
l=\rho-\frac{a^2}{4\rho}.
\label{eq:2:9}
\end{equation}
The Jacobian is positive everywhere in the chart,
\begin{equation}
\frac{\dd l}{\dd\rho}=1+\frac{a^2}{4\rho^2}>0,
\label{eq:2:10}
\end{equation}
and $l(\rho)\to-\infty$ as $\rho\to0^+$ while $l(\rho)\to+\infty$ as $\rho\to\infty$. Hence the map is a smooth bijection $(0,\infty)\to\R$; $\rho=0$ is not a point in the chart, but the limiting representation of the second asymptotic end.

The identity
\begin{equation}
l^2+a^2=\left(\rho+\frac{a^2}{4\rho}\right)^2
\label{eq:2:11}
\end{equation}
fixes the positive conformal factor uniquely:
\begin{equation}
n(\rho)=1+\frac{a^2}{4\rho^2}.
\label{eq:2:12}
\end{equation}
Since $\dd l/\dd\rho=n$ identically, the radial matching condition is satisfied as well. The full metric becomes
\begin{equation}
\dd s^2=-c^2\dd t^2+n^2(\rho)
\left(\dd\rho^2+\rho^2\dd\Omega^2\right).
\label{eq:2:13}
\end{equation}
The throat maps to $\rho_{\rm th}=a/2$, where $n=2$. At the positive end $n\to1$; at the negative end $n\sim a^2/(4\rho^2)$.

The map also makes the Ellis inversion symmetry explicit. Under $l\mapsto-l$, Eq.~\eqref{eq:2:9} gives $\rho\mapsto a^2/(4\rho)$, while the throat $\rho=a/2$ is fixed. Thus the two asymptotic ends are exchanged by the inversion, and the puncture at $\rho=0$ is paired with $\rho=\infty$. This is a geometric symmetry of the ideal chart, not a symmetry of a truncated finite device.

The determinant in the isotropic chart is
\begin{equation}
g_\rho=-c^2n^6\rho^4\sin^2\theta,
\qquad
\sqrt{-g_\rho}=cn^3\rho^2\sin\theta.
\label{eq:2:14}
\end{equation}
For the transformation $(t,\rho,\theta,\phi)\mapsto(t,l,\theta,\phi)$, the four-dimensional Jacobian determinant is $J=\dd l/\dd\rho=n$. Consequently,
\begin{equation}
g_\rho=g_lJ^2,\qquad
\sqrt{-g_\rho}=\sqrt{-g_l}\,|J|,
\label{eq:2:15}
\end{equation}
as follows directly from $\Rcal=n\rho$. 


\section{Constitutive correspondence and exact Maxwell system}
\label{sec:maxwell}

The isotropic metric is not, by itself, a material specification. To obtain the flat-space constitutive tensors, introduce Cartesian laboratory coordinates
\begin{equation}
x=\rho\sin\theta\cos\phi,\quad
y=\rho\sin\theta\sin\phi,\quad
z=\rho\cos\theta .
\label{eq:3:1}
\end{equation}
Then $\dd\rho^2+\rho^2\dd\Omega^2=\delta_{ij}\dd x^i\dd x^j$. In coordinates $x^0=ct$, the metric reads
\begin{equation}
g_{\mu\nu}=\operatorname{diag}(-1,n^2,n^2,n^2),
\quad \sqrt{-g}=n^3,\quad
g^{ij}=n^{-2}\delta^{ij}.
\label{eq:3:2}
\end{equation}
For a static metric with $g_{0i}=0$, the Plebański constitutive correspondence maps the vacuum Maxwell equations on the curved background to a medium in flat space with
\begin{equation}
\frac{\varepsilon_{\rm eff}^{ij}}{\varepsilon_0}
=\frac{\mu_{\rm eff}^{ij}}{\mu_0}
=-\frac{\sqrt{-g}}{g_{00}}g^{ij}.
\label{eq:3:3}
\end{equation}
This is the curved-spacetime constitutive relation derived by Plebański \cite{Plebanski1960}.
The coordinate-design interpretation was developed for electromagnetic field control by Pendry, Schurig, and Smith~\cite{Pendry2006}.

Substitution of Eq.~\eqref{eq:3:2} into Eq.~\eqref{eq:3:3} gives the exact relative constitutive tensors
\begin{equation}
\epsr(\rho)=\mur(\rho)=n(\rho)\mathbf I,
\qquad n(\rho)=1+\frac{a^2}{4\rho^2}.
\label{eq:3:4}
\end{equation}
The medium is isotropic and impedance matched: its local impedance is $Z_0=\sqrt{\mu_0/\varepsilon_0}$ and its local index is $\sqrt{\varepsilon_r\mu_r}=n$. There is no magnetoelectric coupling, and this follows exactly from the static, diagonal form of the metric. The tensors in Eq.~\eqref{eq:3:4} are written in an orthonormal Cartesian laboratory frame; in spherical components they acquire the usual basis factors.

For time-harmonic fields with the convention \(e^{-i\omega t}\), the source-free Maxwell equations in a linear isotropic medium take their standard form~\cite{Jackson1999} (see also Ref.~\cite{Chew1995} for inhomogeneous media). With \(\varepsilon(\rho)=\varepsilon_0 n(\rho)\) and \(\mu(\rho)=\mu_0 n(\rho)\) they read
\begin{align}
\nabla\times\mathbf E
&=i\omega\mu_0 n(\rho)\mathbf H,
\label{eq:3:5}\\
\nabla\times\mathbf H
&=-i\omega\varepsilon_0 n(\rho)\mathbf E,
\label{eq:3:6}\\
\nabla\cdot[n(\rho)\mathbf E]
&=0,\quad
\nabla\cdot[n(\rho)\mathbf H]=0.
\label{eq:3:7}
\end{align}
Eliminating the magnetic field yields the exact vector wave equation
\begin{equation}
\nabla\times\left[
\frac{1}{n(\rho)}
\nabla\times\mathbf E
\right]
=\frac{\omega^2}{c^2}n(\rho)\mathbf E,
\quad
c^{-2}=\varepsilon_0\mu_0.
\label{eq:3:8}
\end{equation}
Equation~\eqref{eq:3:8} holds for any nonvanishing, sufficiently regular index profile. Because the factor \(1/n(\rho)\) stays inside the curl, all gradient-index terms are kept exactly.
Together with the divergence constraints~\eqref{eq:3:7}, this operator gives the complete full-wave description of the medium; numerical methods for such vector problems~\cite{TafloveHagness2005} and the modeling of engineered media~\cite{Tretyakov2003} are well documented. In a three-dimensional inhomogeneous medium, Eq.~\eqref{eq:3:8} cannot in general be replaced by a scalar Helmholtz equation with an $n^2$ factor, because the gradients of the profile couple to the vector constraints. The equivalence we obtain is therefore an equivalence of the full Maxwell equations, not just of ray trajectories.

The law~\eqref{eq:3:4} is an exact continuum target. Any finite passive realization will be dispersive, and its scattered fields will depend on the actual material response and interfaces. Finite embedded transformations can be designed to control reflection in particular geometries~\cite{Rahm2008}, but this does not make an arbitrary truncation reflectionless. Ideal-lens designs have produced optical-wormhole and cloaking configurations~\cite{Courtial2018RayOpticalTO,Belin2019IdealLensCloaks}, including a paraxial shifty cloak~\cite{Courtial2024ShiftyCloaks}; these belong to a different design class and do not imply full-wave equivalence for the profile derived here. For the same reason, the Lorentz model of Sec.~\ref{sec:finite} is judged by its constitutive error, and we draw no conclusion about scattering from that error alone.

\section{Optical interpretation of the puncture}
\label{sec:puncture}

For the ultrastatic metric, the optical spatial line element is $\dd\sigma_{\rm opt}^2=n^2(\rho)(\dd\rho^2+\rho^2\dd\Omega^2)$. The optical radial distance from any fixed $\rho>0$ to the puncture diverges:
\begin{equation}
\begin{aligned}
\int_\rho^{\rho_*}n(s)\,\dd s
&=(\rho_*-\rho)+\frac{a^2}{4}\left(\frac{1}{\rho}-\frac{1}{\rho_*}\right)\\
&\longrightarrow\infty\qquad(\rho\to0^+).
\end{aligned}
\label{eq:4:1}
\end{equation}
With $\bar\rho=a^2/(4\rho)$, the optical metric approaches $\dd\bar\rho^2+\bar\rho^2\dd\Omega^2$ as $\rho\to0^+$. The puncture is therefore the second asymptotically flat end, seen from the laboratory, and not a boundary at finite distance. Filling it in or cutting off the divergent profile removes this end and changes the scattering problem.

\begin{figure*}[!t]
\centering
\includegraphics[width=\textwidth]{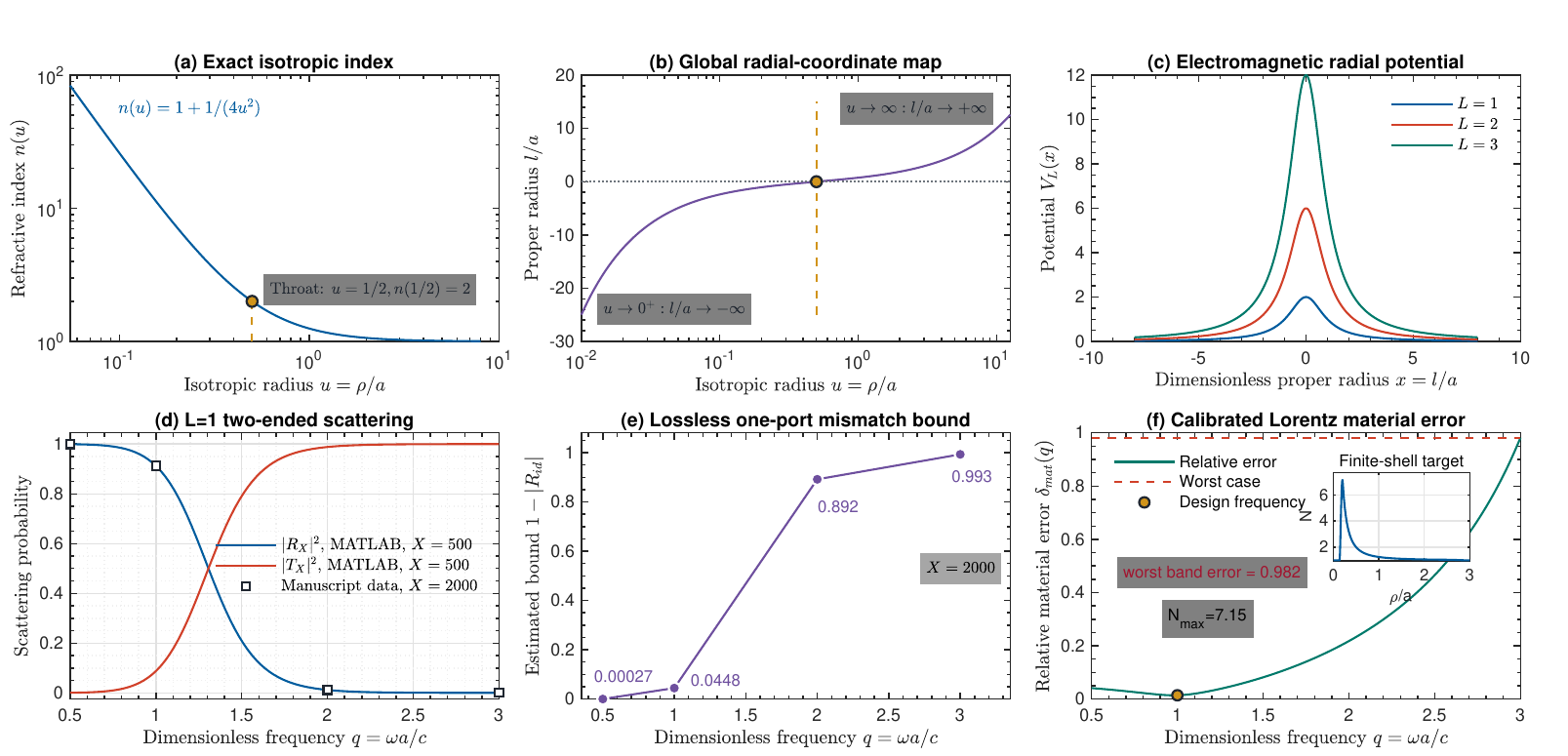}
\caption{(a) exact isotropic index $n(\rho)=1+a^2/(4\rho^2)$; (b) global radial map $l/a=u-1/(4u)$, with $u=\rho/a$; (c) dimensionless potential $V_L(x)=L(L+1)/(1+x^2)$, where $x=l/a$; (d) $L=1$ scattering probabilities versus $q=\omega a/c$; (e) single-port reflection-mismatch bound inferred from those estimates; and (f) relative constitutive error of the representative calibrated Lorentz model, with its finite-shell target shown inset. The throat is at $u=1/2$, where $n=2$. Curves in (d) use $X=500$; markers use $X=2000$ at $q=0.5,1,2,3$. The normalized design uses $a=c=1$, shell radii $(0.12,0.22,2.50,3.00)a$, and Lorentz parameters $(\omega_-,\omega_c,\omega_+,\Omega_0,\gamma)=(0.50,1.00,3.00,4.00,0.25)c/a$. These are representative choices. At the four benchmark frequencies, the finite-$X$ values in (d) and (e) differ from the converged data of Table~\ref{tab:scattering_benchmark} by less than $10^{-6}$ in $|\mathcal R|^2$, far below plotting resolution. Constitutive error alone does not establish finite-device scattering equivalence.}
\label{fig:main-results}
\end{figure*}

\section{Electromagnetic radial operator and two-ended scattering}
\label{sec:scattering}

Rays alone do not determine the Maxwell response at finite frequency. Electromagnetic scattering by the Ellis geometry was analyzed by Cl\'ement~\cite{Clement1984}; a related one-dimensional treatment for static wormholes was given by P\'erez Bergliaffa and Hibberd~\cite{PerezBergliaffaHibberd2000}; and transmission coefficients of electromagnetic waves for a family of wormhole shapes that includes the Ellis case were obtained with higher-order WKB by Konoplya and Zhidenko~\cite{KonoplyaZhidenko2010}. Scalar diffraction and lensing by Ellis wormholes have also been studied~\cite{NambuNodaSakai2019,YooHaradaTsukamoto2013}, but with different field equations and observables. Here we give a self-contained derivation of the electromagnetic radial operator and its spectrum, define the two-ended scattering amplitudes, and evaluate them accurately. We write the areal radius as $R(l)=\sqrt{l^2+a^2}$, so that $\dd s^2=-c^2\dd t^2+\dd l^2+R^2(l)\dd\Omega^2$.

We separate variables in vector spherical harmonics~\cite{ArfkenWeberHarris2013,MorseFeshbach1953}. Let $Y_{Lm}(\theta,\phi)$ be the scalar harmonics on the unit sphere and define the divergence-free axial harmonics $X_A^{Lm}=\epsilon_A{}^B D_B Y_{Lm}$, where $D_A$ and $\epsilon_{AB}$ are the covariant derivative and area form of the sphere. In the axial sector we take $A_A=\psi_L(t,l)X_A^{Lm}$ and $A_t=A_l=0$. Inserting this into $\nabla_\mu F^{\mu\nu}=0$, with $F_{\mu\nu}=2\partial_{[\mu}A_{\nu]}$, and using $D^B(D_BX_A-D_AX_B)=-L(L+1)X_A$, we obtain the Maxwell master equation for the Ellis geometry~\cite{Clement1984}:
\begin{equation}
\left[-\frac{1}{c^2}\frac{\partial^2}{\partial t^2}+\frac{\partial^2}{\partial l^2}-\frac{L(L+1)}{R^2(l)}\right]\psi_L(t,l)=0.
\label{eq:5:1}
\end{equation}
The polar sector is isospectral in source-free Maxwell theory by electromagnetic duality \cite{HehlObukhov2003}. Taking $\psi_L(t,l)=e^{-i\omega t}\psi_L(l)$ yields the exact radial equation
\begin{equation}
\frac{\dd^2\psi_L}{\dd l^2}+\left[k^2-\frac{L(L+1)}{l^2+a^2}\right]\psi_L=0,
\qquad k=\frac{\omega}{c},\quad L\geq1.
\label{eq:5:2}
\end{equation}
The potential comes directly from Maxwell's equations; it is not borrowed from the scalar problem. A minimally coupled massless scalar field, after its radial function is rescaled by $R(l)$, picks up an extra term $R''/R=a^2/(l^2+a^2)^2$ in its effective potential, and this term is absent here. The same distinction appears in Cl\'ement's analysis~\cite{Clement1984}.

This master variable is the radial part of an electromagnetic Debye-potential representation. In a spherically symmetric, ultrastatic background, the two electromagnetic parity sectors can be reconstructed from scalar Debye potentials whose angular dependence is carried by $Y_{Lm}$ and whose radial functions obey Eq.~\eqref{eq:5:2}; electromagnetic duality interchanges the two reconstructions. The Debye formulation therefore provides a scalar generating representation of the vector field, but it does not replace the vector Maxwell system of Sec.~\ref{sec:maxwell}: the physical fields are obtained by angular and radial differential operations on the potentials, and the $L\geq1$ restriction excludes the non-radiative monopole sector.

Introduce $x=l/a$, $q=ka$, and $\lambda_L=L(L+1)$. Equation~\eqref{eq:5:2} becomes
\begin{equation}
\begin{aligned}
H_L\psi_L&=q^2\psi_L,\\
H_L&=-\frac{\dd^2}{\dd x^2}+V_L(x),\qquad
V_L(x)=\frac{\lambda_L}{1+x^2}.
\end{aligned}
\label{eq:5:3}
\end{equation}
The potential is real, smooth, nonnegative, and lies in $L^1(\mathbb R)\cap L^\infty(\mathbb R)$. Since it is bounded, $H_L$ is self-adjoint on $L^2(\mathbb R)$ with domain $H^2(\mathbb R)$~\cite{ReedSimon1979}.

There are no bound states. The argument is the standard one for a nonnegative potential~\cite{DeiftTrubowitz1979}: if $q^2=-\chi^2<0$ and $\psi_L\in H^2(\mathbb R)$ were a nonzero eigenfunction, then
\begin{align}
-\chi^2\|\psi_L\|_2^2&=\langle\psi_L,H_L\psi_L\rangle\\
&=\int_{-\infty}^{\infty}\left(|\psi_L'|^2+V_L|\psi_L|^2\right)\dd x\geq0,
\label{eq:5:4}
\end{align}
which is impossible. At $q=0$ the same identity forces $\psi_L'=0$, and the only constant in $L^2(\mathbb R)$ is zero. For $q>0$, solutions oscillate at both ends and cannot be square integrable~\cite{DeiftTrubowitz1979}. Since $V_L\to0$ at both ends, the essential spectrum is $[0,\infty)$, and therefore
\begin{equation}
\sigma(H_L)=[0,\infty),\qquad \sigma_{\mathrm p}(H_L)=\varnothing,\qquad L\geq1.
\label{eq:5:5}
\end{equation}
This rules out normalizable bound states, but not reflection, transmission, or resonances defined by analytic continuation of the scattering data~\cite{DeiftTrubowitz1979}. The numerical scattering data are given in Table~\ref{tab:scattering_benchmark}, and their dependence on $L$ enters through the centrifugal term in Eq.~\eqref{eq:5:3}.

We build the scattering amplitudes from two solutions of Eq.~\eqref{eq:5:3}. For $q>0$, let $f_+(x,q)$ be the solution that behaves as a unit outgoing wave at the right end and $f_-(x,q)$ the corresponding solution at the left end (the right and left Jost solutions). Both satisfy
\begin{equation}
f_\pm''(x,q)+q^2 f_\pm(x,q)=V_L(x)f_\pm(x,q).
\label{eq:jost-ode}
\end{equation}
Writing the potential term as a source and using the outgoing Green functions of the free equation, $G_+(x,y;q)=\sin[q(y-x)]/q$ for $y\geq x$ and $G_-(x,y;q)=\sin[q(x-y)]/q$ for $y\leq x$, turns Eq.~\eqref{eq:jost-ode} into the Volterra equations below. Their limits of integration fix the asymptotic normalization; they are exact integral representations of the Jost solutions, not closed-form evaluations. Since $V_L$ is integrable on the real line, the standard Jost construction applies for real $q>0$~\cite{DeiftTrubowitz1979,ChadanSabatier1989,Yafaev1992,Taylor1972}:
\begin{align}
f_+(x,q)&=e^{iqx}+\int_x^{\infty}\frac{\sin[q(y-x)]}{q}V_L(y)f_+(y,q)\,\dd y,
\label{eq:5:6}\\
f_-(x,q)&=e^{-iqx}+\int_{-\infty}^{x}\frac{\sin[q(x-y)]}{q}V_L(y)f_-(y,q)\,\dd y.
\label{eq:5:7}
\end{align}
Thus $f_+\sim e^{iqx}$ as $x\to+\infty$ and $f_-\sim e^{-iqx}$ as $x\to-\infty$. Neither is by itself the physical state with a given incident wave; that state is obtained by normalizing the incident part of $f_+$, as shown below. We define the Wronskian $W[u,v]=uv'-u'v$, which is independent of $x$ for any two solutions because Eq.~\eqref{eq:5:3} has no first-derivative term.

For a wave incident from the $l=-\infty$ end, the standard one-dimensional scattering normalization~\cite{DeiftTrubowitz1979} is
\begin{equation}
\psi_L(x)\sim\begin{cases}
e^{iqx}+\mathcal R_L(q)e^{-iqx},&x\to-\infty,\\
\mathcal T_L(q)e^{iqx},&x\to+\infty.
\end{cases}
\label{eq:5:8}
\end{equation}
At the left end we write $f_+(x,q)\sim A_L(q)e^{iqx}+B_L(q)e^{-iqx}$. For real $q>0$, $W[f_-,f_+]$ cannot vanish, since otherwise a solution would be outgoing at both ends, which violates current conservation for a real potential~\cite{DeiftTrubowitz1979}. Evaluating the Wronskians gives
\begin{equation}
A_L(q)=\frac{W[f_-,f_+]}{2iq},\qquad B_L(q)=\frac{W[f_+,f_-^*]}{2iq}.
\label{eq:5:9}
\end{equation}
It follows that the reflection and transmission amplitudes are
\begin{equation}
\mathcal T_L(q)=\frac{2iq}{W[f_-,f_+]},\qquad
\mathcal R_L(q)=\frac{W[f_+,f_-^*]}{W[f_-,f_+]}.
\label{eq:5:10}
\end{equation}
Equations~\eqref{eq:5:6}--\eqref{eq:5:10} are exact and involve no geometrical-optics approximation. We do not claim that the amplitudes reduce to elementary or known special functions; they are evaluated numerically and asymptotically in Sec.~\ref{sec:numerics}.

For real $q$, the conserved radial current used in one-dimensional scattering theory~\cite{DeiftTrubowitz1979} is
\begin{equation}
j_L=\frac{1}{2i}\left(\psi_L^*\psi_L'-\psi_L\psi_L'^*\right).
\label{eq:5:11}
\end{equation}
Evaluating this current at both ends with the normalization~\eqref{eq:5:8} gives the flux-conservation identity~\cite{DeiftTrubowitz1979}
\begin{equation}
|\mathcal R_L(q)|^2+|\mathcal T_L(q)|^2=1,\qquad q>0.
\label{eq:5:12}
\end{equation}
Everything in this section refers to the ideal, lossless, two-ended geometry. A finite, truncated, dispersive device has its own scattering matrix, set by its material response, interfaces, and boundary conditions.

\subsection{Numerical evaluation of the scattering data}
\label{sec:numerics}

\begin{figure*}[t]
\centering
\includegraphics[width=0.9\textwidth]{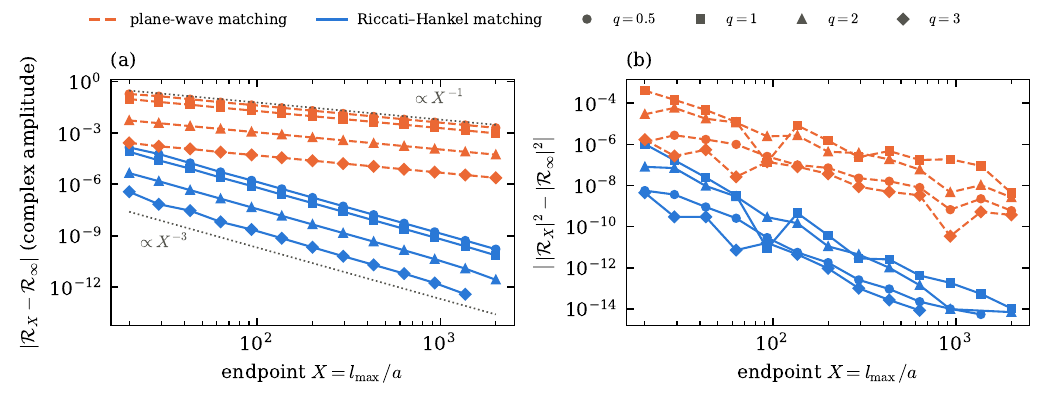}
\caption{Convergence of the $L=1$ reflection data with the endpoint $X$ ($20\le X\le2000$) for $q=0.5,1,2,3$. (a) Complex amplitude; (b) reflection probability. Dashed lines: plane-wave matching; solid lines: Riccati--Hankel matching, Eq.~\eqref{eq:num:rh}, with integration tolerances $3\times10^{-14}$. The reference $\mathcal R_\infty=(8\mathcal R_{4000}-\mathcal R_{2000})/7$ is the Richardson extrapolation of Riccati--Hankel results at $X=2000$ and $4000$ computed with the same tolerances; it lies outside the plotted range and is accurate to about $3\times10^{-11}$. Differences below $10^{-13}$ are not shown in (a).}
\label{fig:convergence}
\end{figure*}

\paragraph*{Asymptotic matching.}
For $|x|\gg1$ the potential in Eq.~\eqref{eq:5:3} has the expansion
\begin{equation}
V_L(x)=\frac{\lambda_L}{x^2}-\frac{\lambda_L}{x^4}+O(x^{-6}),
\label{eq:num:tail}
\end{equation}
so its leading tail is exactly the flat-space centrifugal barrier. If the numerical solution is matched to plane waves at a finite endpoint $X$, this $O(\lambda_L/x^2)$ tail is ignored and the complex amplitudes carry an $O(\lambda_L/qX)$ error, almost all of it in the phase. We avoid this by imposing the outgoing condition with Riccati-Hankel functions,
\begin{equation}
\begin{aligned}
&f_+(X,q)=\hat h^{(1)}_L(qX),\quad
f_+'(X,q)=q\,\hat h^{(1)\prime}_L(qX),\\
&\hat h^{(1,2)}_L(z)=z\,h^{(1,2)}_L(z),
\end{aligned}
\label{eq:num:rh}
\end{equation}
and decompose the propagated solution at $x=-X$ as $f_+=\alpha\,\hat h^{(2)}_L(q|x|)+\beta\,\hat h^{(1)}_L(q|x|)$. Since $\hat h^{(1,2)}_L(z)\to(\mp i)^{L+1}e^{\pm iz}$, the amplitudes defined by Eq.~\eqref{eq:5:8} are
\begin{equation}
\mathcal T_L=\frac{(-1)^{L+1}}{\alpha},\qquad
\mathcal R_L=(-1)^{L+1}\frac{\beta}{\alpha}.
\label{eq:num:RT}
\end{equation}
The neglected part of the tail is now $O(\lambda_L x^{-4})$, and the truncation error falls as $X^{-3}$. Figure~\ref{fig:convergence} confirms both rates. With plane-wave matching the complex reflection amplitude converges only as $X^{-1}$, with an error of order $10^{-3}$ at $X=2000$; with Riccati--Hankel matching it converges as $X^{-3}$, and at $X\simeq10^3$ the error is below $2\times10^{-9}$ at all four benchmark frequencies. The reflection probability converges faster in both cases because the leading error is a phase. For the same reason, the flux identity cannot detect it: a pure phase error leaves $|\mathcal R|^2+|\mathcal T|^2$ unchanged.

We integrate with the DOP853 scheme~\cite{HairerNorsettWanner1993} at relative and absolute tolerances of $10^{-12}$ and $10^{-14}$. To check the integrator independently, we repeated the $L=1$ calculation at $X=60$ with a 25-digit Taylor-series solver; the reflection probabilities agree to $8\times10^{-14}$ at all four benchmark frequencies. Table~\ref{tab:scattering_benchmark} lists the results at $X=1000$. To estimate their accuracy we use a reference value that does not depend on the plotted range: Riccati--Hankel results at $X=2000$ and $4000$, computed with tighter tolerances ($3\times10^{-14}$) and combined by Richardson extrapolation for the $X^{-3}$ error. The tighter tolerances are needed because, at $10^{-12}$, the integration error accumulated over several hundred wavelengths reaches about $10^{-10}$ at $X\simeq10^3$, comparable to the truncation error itself. Against this reference, which is accurate to about $3\times10^{-11}$, the entries of Table~\ref{tab:scattering_benchmark} are accurate to $1.4\times10^{-9}$ in the complex amplitude and to $5\times10^{-14}$ in $|\mathcal R|^2$. These are numerical error estimates rather than rigorous bounds.

\begin{table*}[t]
\caption{Electromagnetic $L=1$ scattering data for the ideal Ellis geometry, obtained with Riccati-Hankel matching at $X=1000$. Complex amplitudes are accurate to about $2\times10^{-9}$ and reflection probabilities to about $10^{-12}$ (see text and Fig.~\ref{fig:convergence}); the last column is the single-port mismatch bound of Eq.~\eqref{eq:6:4}.}
\label{tab:scattering_benchmark}
\setlength{\tabcolsep}{3pt}\small
\begin{tabular}{cccc}
\toprule
$q$ & $\mathcal R_1$ & $|\mathcal R_1|^2$ & $1-|\mathcal R_1|$\\
\midrule
0.5 & $\phantom{-}0.97879535+0.20351740\,i$ & $0.999459662095$ & $2.70205458\times10^{-4}$\\
1.0 & $\phantom{-}0.57797220+0.76051002\,i$ & $0.912427355366$ & $4.47893660\times10^{-2}$\\
2.0 & $-0.10742029+0.01447842\,i$ & $1.17487429277\times10^{-2}$ & $0.891608382$\\
3.0 & $-0.00634242-0.00339160\,i$ & $5.17292536\times10^{-5}$ & $0.992807695$\\
\bottomrule
\end{tabular}
\end{table*}

\begin{figure*}[t]
\centering
\includegraphics[width=0.9\textwidth]{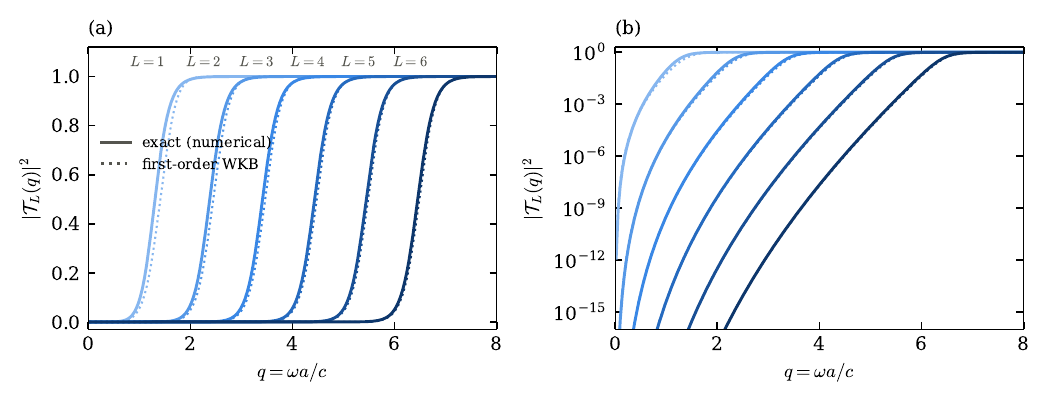}
\caption{Transmission probabilities $|\mathcal T_L(q)|^2$ for $L=1,\dots,6$ on (a) linear and (b) logarithmic scales. Solid lines: numerical; dotted lines: first-order WKB.}
\label{fig:greybody}
\end{figure*}

\begin{figure*}[t]
\centering
\includegraphics[width=\textwidth]{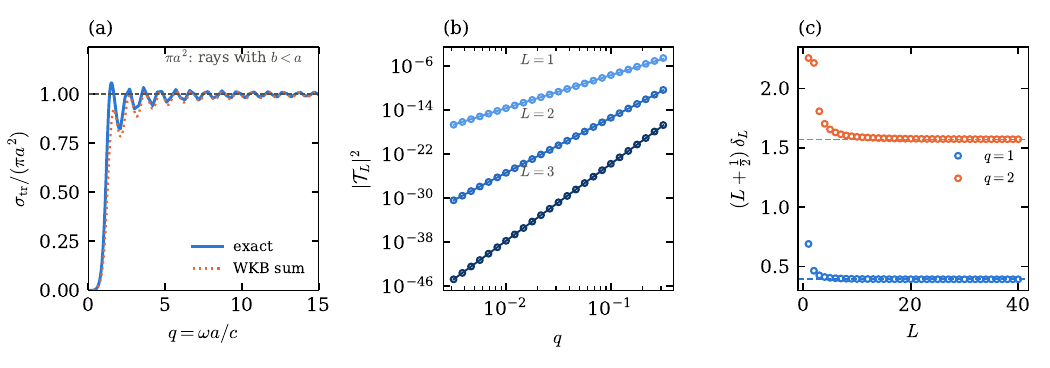}
\caption{(a) Transmission cross section, Eq.~\eqref{eq:num:sigma}, compared with the WKB partial-wave sum and with the geometric value $\pi a^2$. (b) Low-frequency transmission (symbols) and the closed form of Eq.~\eqref{eq:num:lowq} (lines). (c) Large-$L$ phase shifts compared with Eq.~\eqref{eq:num:delta} (dashed).}
\label{fig:cross}
\end{figure*}

\paragraph*{Partial waves and low-frequency law.}
Grey-body factors of electromagnetic fields in Morris--Thorne wormholes, the Ellis shape included, were first computed with higher-order WKB formulas in Ref.~\cite{KonoplyaZhidenko2010}; methods and results are reviewed in Ref.~\cite{KonoplyaZhidenko2026Review}. Figure~\ref{fig:greybody} displays the transmission probabilities $|\mathcal T_L(q)|^2$ obtained here by direct integration for $L=1,\ldots,6$. The transmission edge, defined by $|\mathcal T_L|^2=1/2$, occurs near the barrier maximum: the computed values are $q=1.30$, $2.39$, and $3.43$ for $L=1$, $2$, and $3$, respectively, and follow the approximate trend $q\simeq L+0.3$. First-order WKB, using the Kemble expression below the barrier top and the parabolic-barrier approximation above it, captures the location of the edge but is not quantitatively reliable for the lowest multipoles. Its maximum absolute error in $|\mathcal T_L|^2$ decreases from $0.18$ at $L=1$ to $0.045$ at $L=6$, approximately as $1/L$. Higher-order WKB formulas are considerably more accurate~\cite{KonoplyaZhidenko2010,KonoplyaZhidenko2026Review}; the first-order comparison is shown only to indicate where the semiclassical picture applies, and all quantitative results below come from direct numerical integration. In the low-frequency limit, the zero-energy equation $(1+x^2)\psi^{\prime\prime}=\lambda_L\psi$ can be represented using Gegenbauer functions of order $-\tfrac12$ with argument $ix$. Matching the zero-energy solutions to Riccati-Bessel waves at both asymptotic ends yields the threshold scaling $|\mathcal T_L|^2\propto q^{4L+2}$. For $L=1,2,3,4$, the corresponding matching coefficient is given below and has been checked independently against high-precision numerical solutions:
\begin{equation}
|\mathcal T_L(q)|^2\simeq
\frac{4\,[(L-1)!\,(L+1)!]^2\,q^{4L+2}}{\pi^2\,[(2L-1)!!\,(2L+1)!!]^4}
\quad (q\to0).
\label{eq:num:lowq}
\end{equation}
For $L=1$, Eq.~\eqref{eq:num:lowq} reduces to $|\mathcal T_1|^2\simeq16q^6/(81\pi^2)$. For each of the four tested multipoles, the coefficient agrees with the high-precision numerical solution; at $q=10^{-3}$ the relative discrepancy is below $10^{-4}$ (see Fig.~\ref{fig:cross}(b)). We therefore regard the displayed coefficient as numerically validated for $L=1,2,3,4$. The matching calculation establishes the stated threshold exponent; the coefficient validation is reported only for the multipoles explicitly tested.

\paragraph*{Observable cross sections.}
For an unpolarized plane wave incident from one asymptotic end, the power transmitted through the throat defines the cross section
\begin{equation}
\sigma_{\rm tr}(\omega)=\frac{\pi}{k^2}\sum_{L\geq1}(2L+1)\,|\mathcal T_L(q)|^2,
\label{eq:num:sigma}
\end{equation}
where both electromagnetic parities contribute equally because they are isospectral. As Fig.~\ref{fig:cross}(a) shows, $\sigma_{\rm tr}$ grows as $q^4$ at low frequency and then approaches the geometric-optics value $\pi a^2$, the cross section of the rays with impact parameter $b<a$ that pass through the throat; at $q=15$ it equals $0.997\,\pi a^2$. On top of this rise there are damped oscillations about $\pi a^2$, one for each partial wave that opens, similar to those in the absorption cross section of black holes. They are a pure wave effect with no counterpart in ray optics. Oscillatory approaches to geometric-optics cross sections are familiar in black-hole wave scattering, where diffraction and unstable-ray dynamics control the high-frequency structure~\cite{Sanchez1978Absorption,DecaniniEspositoFareseFolacci2011}. Here the oscillations arise in the throat-transmission observable and are computed directly from the Ellis electromagnetic partial waves.
The total cross section for scattering back into the incident end behaves very differently. At large $L$ the $-\lambda_L x^{-4}$ tail gives the phase shifts
\begin{equation}
\delta_L\simeq\frac{\pi q^2}{8\,(L+\tfrac12)},\qquad L\gg q,
\label{eq:num:delta}
\end{equation}
as Fig.~\ref{fig:cross}(c) confirms. The semiclassical deflection angle $\Theta=2\,\partial\delta_L/\partial L$ then reproduces the known weak-field result for the Ellis wormhole, $|\Theta|=\pi a^2/(4b^2)$. Since $\delta_L$ falls off only as $1/L$, the sum $\sum_L(2L+1)\sin^2\delta_L$ diverges logarithmically. In the laboratory this is the familiar forward-peak divergence produced by an index tail $n^2-1\simeq a^2/(2\rho^2)$: only $\sigma_{\rm tr}$ and the differential cross section at nonzero angle are finite. A finite device has no such tail beyond $\rho_{\max}$, so its total cross section is finite. This gives a second far-field signature, independent of the bound in Sec.~\ref{sec:finite}, that separates any finite device from the ideal profile.

\section{Finite-device truncation and the scope of equivalence}
\label{sec:finite}

The exact continuum representation occupies the punctured chart $\R^3\setminus\{0\}$ and has an unbounded constitutive profile as $\rho\to0^+$. If a physical device excludes $\rho<\rho_{\min}$, the ideal index on the retained domain is bounded above by
\begin{equation}
n_{\max}=1+\frac{a^2}{4\rho_{\min}^2}.
\label{eq:6:1}
\end{equation}
However small $\rho_{\min}$ is, it removes the region $\bar\rho=a^2/(4\rho)\to\infty$ and with it the second optical end. This is not a small local change: the omitted interval has infinite optical length, since $\int_0^{\rho_{\min}}n(\rho)\,\dd\rho=\infty$.

A finite device also needs an outer radius $\rho_{\max}$ and some interface or termination. Because $\varepsilon_r=\mu_r$, the local wave impedance is $Z_0$ and a planar interface at normal incidence is matched. A finite spherical profile still scatters, however, because radial inhomogeneity and angular-momentum matching enter the full wave problem~\cite{Jackson1999}. Reflectionless finite transformations exist for particular constructions~\cite{Rahm2008}, but the termination remains part of the device and has to be modeled.

\subsection{A single-port obstruction to lossless finite-device equivalence}
The difference between the ideal geometry and a regular finite device can be made precise without specifying the cutoff. Fix a real frequency $\omega>0$ and one decoupled partial-wave channel, and let $\mathcal R_{\mathrm{id},L}(\omega)$ and $\mathcal T_{\mathrm{id},L}(\omega)$ be the ideal reflection and transmission amplitudes, normalized to unit incoming flux. For a real, integrable potential, Eq.~\eqref{eq:5:10} gives a finite, nonzero transmission amplitude at every $q=\omega a/c>0$, so Eq.~\eqref{eq:5:12} implies
\begin{equation}
\begin{aligned}
 |\mathcal R_{\mathrm{id},L}|^2+|\mathcal T_{\mathrm{id},L}|^2&=1,\\
 0<|\mathcal T_{\mathrm{id},L}|&\leq1,
 \qquad |\mathcal R_{\mathrm{id},L}|<1.
\end{aligned}
\label{eq:6:2}
\end{equation}
Now consider a finite, spherically symmetric, regular realization that is lossless, has vacuum outside a bounded region, and couples this partial wave to no other propagating port. Its exterior problem then has one incoming and one outgoing channel. Since $V_L(x)$ is even, the ideal reflection amplitudes for incidence from either end have the same modulus, and they coincide when the reference planes are placed symmetrically about the throat. We compare the finite device with the ideal amplitude for its exterior channel. Energy conservation makes the device's one-channel reflection coefficient unimodular,
\begin{equation}
 |\mathcal R_{\mathrm{fin},L}(\omega)|=1.
 \label{eq:6:3}
\end{equation}
whatever the structure of its core and shell. The reverse triangle inequality then gives the mismatch bound
\begin{align}
 \left|\mathcal R_{\mathrm{fin},L}-\mathcal R_{\mathrm{id},L}\right|
 &\geq \left||\mathcal R_{\mathrm{fin},L}|-|\mathcal R_{\mathrm{id},L}|\right|\\
 &=1-|\mathcal R_{\mathrm{id},L}|\\
 &=1-\sqrt{1-|\mathcal T_{\mathrm{id},L}|^2}>0.
 \label{eq:6:4}
\end{align}
The phases of the two amplitudes must refer to the same exterior reference plane, but the bound itself does not depend on them. Specifying the input and output channels is equally central in general scattering-matrix treatments, for example of reflectionless scattering modes~\cite{Sweeney2020ReflectionlessModes}. Equation~\eqref{eq:6:4} is a statement about channel structure. It says nothing about constitutive error and does not apply to every finite medium. A lossy one-port can have $|\mathcal R|<1$ because it absorbs power, although absorption is physically different from transmission into a second asymptotic end. Likewise, a device with an explicit auxiliary port has a larger unitary scattering matrix and must be compared through a channel map that includes that port. The theorem therefore excludes exact matching of the complex reflection amplitude by lossless one-port devices wherever the ideal transmission is nonzero, but not approximate matching by lossy or multiport designs.

For the $L=1$ channel, the last column of Table~\ref{tab:scattering_benchmark} evaluates the bound in Eq.~\eqref{eq:6:4} from the converged reflection amplitudes: it grows from $2.70\times10^{-4}$ at $q=0.5$, where the ideal geometry is almost perfectly reflecting, to $0.993$ at $q=3$, where it is almost perfectly transmitting. At high frequency, a lossless single-port device therefore cannot reproduce even the order of magnitude of the ideal reflection amplitude.

\subsection{Causal dispersive regularization and finite-band error}

The ideal target is the frequency-independent profile of Eq.~\eqref{eq:3:4}, obtained from the Pleba\'nski correspondence~\cite{Plebanski1960}. No real material follows it at all frequencies, since causal linear response requires the right high-frequency behavior~\cite{LandauLifshitz}; see also Refs.~\cite{Brillouin1960,Milton2002,Nussenzveig1972}.

Choose radii ordered as
\begin{equation}
0<\rho_{\min}<\rho_1<\rho_2<\rho_{\max}.
\label{eq:6:5}
\end{equation}
Define the smooth step
\begin{align}
p(x)&=\begin{cases}0,&x\leq0,\\ \exp(-1/x),&x>0,\end{cases}\\
\eta(x)&=\frac{p(x)}{p(x)+p(1-x)}.
\label{eq:6:6}
\end{align}
The denominator is strictly positive for all real $x$, so $\eta$ is smooth, with $\eta(x)=0$ for $x\leq0$ and $\eta(x)=1$ for $x\geq1$. The spatial window
\begin{equation}
S(\rho)=\eta\!\left(\frac{\rho-\rho_{\min}}{\rho_1-\rho_{\min}}\right)
\left[1-\eta\!\left(\frac{\rho-\rho_2}{\rho_{\max}-\rho_2}\right)\right]
\label{eq:6:7}
\end{equation}
satisfies $0\leq S\leq1$, vanishes for $\rho\leq\rho_{\min}$ and $\rho\geq\rho_{\max}$, and equals one on $[\rho_1,\rho_2]$. It is a regularization introduced for the finite device and is not meant to preserve the ideal scattering matrix. The resulting target is
\begin{equation}
N(\rho)=1+S(\rho)\frac{a^2}{4\rho^2}.
\label{eq:6:8}
\end{equation}
Thus $N(\rho)\geq1$, it equals vacuum outside the finite shell, and it is smooth and bounded. Since $0\leq S(\rho)\leq1$ and $\rho\geq\rho_{\min}$ wherever the nonzero profile is retained, its maximum obeys the direct bound
\begin{equation}
1\leq N_{\max}:=\max_{\rho>0}N(\rho)
\leq1+\frac{a^2}{4\rho_{\min}^2}.
\label{eq:6:9}
\end{equation}
Because $S$ vanishes near the origin, Eq.~\eqref{eq:6:8} extends smoothly to $\rho=0$ with $N(0)=1$. The construction thus replaces the inner end by a smooth core and returns to vacuum outside $\rho_{\max}$; it is no longer the global Ellis profile.

Let the operating band be $0<\omega_-\leq\omega\leq\omega_+$, with design frequency $\omega_c\in[\omega_-,\omega_+]$. Choose a Lorentz resonance and damping rate such that
\begin{equation}
\Omega_0>\omega_+,\qquad \gamma>0,
\label{eq:6:10}
\end{equation}
and define
\begin{equation}
\Delta_c=\Omega_0^2-\omega_c^2>0,\qquad
A_c=\frac{\Delta_c^2+\gamma^2\omega_c^2}{\Delta_c}.
\label{eq:6:11}
\end{equation}
Using the damped Lorentz-oscillator form for the constitutive response~\cite{LandauLifshitz}, we calibrate its real part to equal $N(\rho)$ at $\omega_c$. This fixes the dimensionless oscillator strength to
\begin{equation}
F(\rho)=\frac{N(\rho)-1}{\Omega_0^2}
\frac{\Delta_c^2+\gamma^2\omega_c^2}{\Delta_c}.
\label{eq:6:12}
\end{equation}
We use the matched Lorentz model
\begin{align}
\varepsilon_r(\rho,\omega)&=1+\frac{F(\rho)\Omega_0^2}
{\Omega_0^2-\omega^2-\ii\gamma\omega},
\label{eq:6:13}\\
\mu_r(\rho,\omega)&=1+\frac{F(\rho)\Omega_0^2}
{\Omega_0^2-\omega^2-\ii\gamma\omega}.
\label{eq:6:14}
\end{align}
The absolute parameters are $\varepsilon=\varepsilon_0\varepsilon_r$ and $\mu=\mu_0\mu_r$. Since $\varepsilon_r=\mu_r$, the impedance ratio is $Z/Z_0=\sqrt{\mu_r/\varepsilon_r}=1$ on the physical branch~\cite{Jackson1999}. The oscillator strength~\eqref{eq:6:12} is dimensionless. The calibration fixes the real response at $\omega_c$ but cannot remove the loss that comes with $\gamma>0$.

Writing $D(\omega)=\Omega_0^2-\omega^2$, the standard damped Lorentz response~\cite{LandauLifshitz} has the following real and imaginary parts for real positive frequencies:
\begin{align}
\operatorname{Re}\varepsilon_r(\rho,\omega)&=1+
\frac{F(\rho)\Omega_0^2D(\omega)}{D^2(\omega)+\gamma^2\omega^2},
\label{eq:6:15}\\
\operatorname{Im}\varepsilon_r(\rho,\omega)&=
\frac{F(\rho)\Omega_0^2\gamma\omega}{D^2(\omega)+\gamma^2\omega^2}.
\label{eq:6:16}
\end{align}
The same relations hold for $\mu_r$. Substituting Eq.~\eqref{eq:6:12} at $\omega=\omega_c$ gives
\begin{align}
\operatorname{Re}\varepsilon_r(\rho,\omega_c)&=N(\rho),
\label{eq:6:17}\\
\operatorname{Re}\mu_r(\rho,\omega_c)&=N(\rho),
\label{eq:6:18}\\
\operatorname{Im}\varepsilon_r(\rho,\omega_c)&=
\operatorname{Im}\mu_r(\rho,\omega_c)
=[N(\rho)-1]\frac{\gamma\omega_c}{\Delta_c}.
\label{eq:6:19}
\end{align}
The real target is thus reproduced exactly at the design frequency, but wherever $N(\rho)>1$ the complex response differs from it as long as $\gamma>0$: calibration does not remove absorption.

For the time convention $e^{-\ii\omega t}$, the poles of the damped Lorentz response lie in the lower half-plane, consistently with the causal-response convention~\cite{Skaar2006}:
\begin{equation}
\omega_{\mathrm{p},\pm}=-\frac{\ii\gamma}{2}
\pm\sqrt{\Omega_0^2-\frac{\gamma^2}{4}}.
\label{eq:6:20}
\end{equation}
Both poles lie in the lower half-plane for $\Omega_0,\gamma>0$, including the overdamped case. Moreover, $N\geq1$ implies $F\geq0$, so by Eq.~\eqref{eq:6:16} the imaginary parts are nonnegative at positive frequencies. The model is therefore causal and passive in the usual local sense~\cite{LandauLifshitz,Skaar2006}. These are properties of a phenomenological model; we do not claim that a particular material realizes it.

Subtracting the real target from the Lorentz response~\cite{LandauLifshitz} gives the exact complex error
\begin{equation}
\varepsilon_r(\rho,\omega)-N(\rho)=
[N(\rho)-1]\left[\frac{A_c}{D(\omega)-\ii\gamma\omega}-1\right],
\label{eq:6:21}
\end{equation}
and the same identity holds for $\mu_r-N$. Define the worst-case relative constitutive error over the device and operating band by
\begin{equation}
\begin{aligned}
\delta_{\mathrm{mat}}&=\sup_{\substack{\rho\in[\rho_{\min},\rho_{\max}]\\
\omega\in[\omega_-,\omega_+]}}\max\Biggl\{
\frac{|\varepsilon_r(\rho,\omega)-N(\rho)|}{N(\rho)},\\[-1pt]
&\hspace{2.8em}\frac{|\mu_r(\rho,\omega)-N(\rho)|}{N(\rho)}\Biggr\}.
\end{aligned}
\label{eq:6:22}
\end{equation}
Because the two constitutive responses are identical and the spatial and frequency factors separate, this reduces exactly to
\begin{equation}
\begin{aligned}
\delta_{\mathrm{mat}}&=\left(1-\frac{1}{N_{\max}}\right)
\sup_{\omega\in[\omega_-,\omega_+]}\\[-2pt]
&\quad\times
\frac{\sqrt{[A_c-D(\omega)]^2+\gamma^2\omega^2}}
{\sqrt{D^2(\omega)+\gamma^2\omega^2}}.
\end{aligned}
\label{eq:6:23}
\end{equation}
This expression is exact, not a small-dispersion approximation, and the remaining maximization over frequency reduces to a finite set of candidates. With $x=\omega^2$, $U=\Omega_0^2$, and $B=\gamma^2$, define
\begin{equation}
G(x)=\frac{(A_c-U+x)^2+Bx}{(U-x)^2+Bx}.
\label{eq:6:24}
\end{equation}
Then
\begin{equation}
\delta_{\mathrm{mat}}=\left(1-\frac{1}{N_{\max}}\right)
\sqrt{\max_{x\in\mathcal C}G(x)},
\label{eq:6:25}
\end{equation}
where $\mathcal C$ contains $x_-=\omega_-^2$, $x_+=\omega_+^2$, and every real stationary point in $[x_-,x_+]$. Since $A_c>0$, the stationary points satisfy
\begin{equation}
x^2+(A_c-2U)x+U^2-A_cU-BU+\frac{A_cB}{2}=0,
\label{eq:6:26}
\end{equation}
so the only possible interior candidates are
\begin{equation}
x_{\mathrm{stat},\pm}=\frac{2U-A_c\pm
\sqrt{A_c^2-2A_cB+4BU}}{2},
\label{eq:6:27}
\end{equation}
provided the discriminant is nonnegative and the roots lie in the band. Evaluating $G$ at the band edges and at any admissible stationary point therefore gives the exact worst-case constitutive error. The finite-device construction and its dispersive error are thus fully specified by Eqs.~\eqref{eq:6:8}--\eqref{eq:6:25}. Fitting an effective response in this way is different from retrieving the parameters of a fabricated metamaterial~\cite{SmithSchultzMarkosSoukoulis2002}.

For the dispersive medium, the source-free equation for the electric field is~\cite{Jackson1999}
\begin{equation}
\nabla\times\left[\frac{1}{\mu_r(\rho,\omega)}
\nabla\times\mathbf E\right]
=\frac{\omega^2}{c^2}\varepsilon_r(\rho,\omega)\mathbf E,
\label{eq:6:28}
\end{equation}
with constraints
\begin{equation}
\nabla\cdot[\varepsilon_r(\rho,\omega)\mathbf E]=0,
\quad
\nabla\cdot[\mu_r(\rho,\omega)\mathbf H]=0.
\label{eq:6:29}
\end{equation}
These equations keep the full spatial dependence of the material functions inside the differential operator, with no scalar or ray reduction~\cite{Jackson1999}.

The regularization changes the global structure of the optical geometry. In the ideal model $\rho\to0^+$ is a second end at infinite optical distance, whereas the regularized profile returns smoothly to vacuum in the core. The two define different scattering problems, and matching the material parameters alone cannot make their scattering matrices equal.

The scattering framework needed for a proper comparison is standard~\cite{ColtonKress2013,Newton1982,Taylor1972,ChadanSabatier1989,Yafaev1992,FelsenMarcuvitz1994}. What is specific here is the mismatch between the two-ended channel space of the ideal geometry and that of a finite device with a regular core. A wave-level comparison has to match the asymptotic channels of the regularized matrix $S_{\mathrm{reg}}(\omega)$ and the ideal matrix $S_{\mathrm{id}}(\omega)$, with consistent normalization, before the two can be compared; matrices of different dimension cannot simply be subtracted. Once the channel spaces are matched and a norm is chosen, the scattering error is~\cite{Newton1982}
\begin{equation}
\delta_S(\omega)=\left\|S_{\mathrm{reg}}(\omega)-S_{\mathrm{id}}(\omega)\right\|.
\label{eq:6:30}
\end{equation}
Unlike the constitutive error~\eqref{eq:6:23}, this quantity cannot be read off the local material law; it requires solving both full-wave problems. In particular, a reflecting inner termination and a continuation that carries an outgoing wave into a second end are physically different. The present analysis therefore gives a complete and quantitative answer to the question posed for the ideal geometry and for the specified finite-device class: continuum equivalence is exact, while exact one-port scattering equivalence is excluded by the channel-mismatch bound. The causal Lorentz model is assessed independently through its constitutive error. These conclusions do not require a device-specific full-wave simulation; such a simulation would quantify approximate performance for a separately specified lossy or multiport implementation, rather than alter the no-go result proved here.

\section{Discussion and conclusions}
\label{sec:conclusions}

The results establish a sequence of distinct statements: an exact coordinate representation of the ideal geometry; its exact continuum constitutive realization; the two-ended electromagnetic scattering problem and its converged multipole solution; a channel-based obstruction to exact matching by a specified finite-device class; and a finite dispersive material benchmark. Figure~\ref{fig:main-results} summarizes these results. It should be read as a summary of the analytical and numerical construction, not as evidence that a fabricated device reproduces the ideal scattering matrix.

Panels~\ref{fig:main-results}(a) and (b) display the global isotropic map. Equation~\eqref{eq:2:9} is a smooth bijection from $\rho>0$ to the full signed radial line, with conformal factor $n(\rho)=1+a^2/(4\rho^2)$. The throat is mapped to $\rho=a/2$, where $n=2$, while $\rho\to\infty$ and $\rho\to0^+$ represent the positive and negative asymptotic ends, respectively. As shown by the optical-distance relation in Eq.~\eqref{eq:4:1}, the puncture lies at infinite optical distance and represents the second asymptotic end; it is not a curvature singularity of the Ellis spacetime. A regular core therefore changes the optical domain rather than simply smoothing a local coordinate feature.

At the continuum level, Eq.~\eqref{eq:3:4} gives the exact constitutive correspondence $\boldsymbol{\varepsilon}_r=\boldsymbol{\mu}_r=n(\rho)\mathbf I$. The coefficient remains inside the curl in Eq.~\eqref{eq:3:8}, so the resulting operator reproduces the source-free Maxwell equations and their constraints on the ideal geometry, rather than only the ray trajectories. Panel~\ref{fig:main-results}(c) shows $V_L(x)=L(L+1)/(1+x^2)$. As established for the radial operator in Eq.~\eqref{eq:5:5}, its nonnegative, decaying potential yields spectrum $[0,\infty)$ with no point spectrum.

Panel~\ref{fig:main-results}(d) shows the $L=1$ reflection and transmission probabilities: reflection dominates at low frequency, while transmission approaches unity above the barrier. Riccati-Hankel matching reduces the finite-endpoint error to $O(X^{-3})$; Table~\ref{tab:scattering_benchmark} reports the converged values and their stated numerical precision, with an independent high-precision integration used as a check. The multipole analysis establishes the threshold exponent and validates the coefficient in Eq.~\eqref{eq:num:lowq} for $L=1,2,3,4$. It also shows that $\sigma_{\rm tr}\to\pi a^2$ at high frequency and that the ideal profile's total scattering cross section has a logarithmic forward divergence. A finite-support profile lacks the long-range tail responsible for this divergence and has a finite total cross section.

The finite-device result is the reflection-mismatch bound~\eqref{eq:6:4}. In the ideal geometry, nonzero transmission into the second end gives $|\mathcal R_{\mathrm{id},L}|<1$, whereas a regular, lossless finite realization with one exterior channel satisfies $|\mathcal R_{\mathrm{fin},L}|=1$. Consequently,
\[
\left|\mathcal R_{\mathrm{fin},L}-\mathcal R_{\mathrm{id},L}\right|
\geq 1-|\mathcal R_{\mathrm{id},L}|>0.
\]
For $L=1$, panel~\ref{fig:main-results}(e) and Table~\ref{tab:scattering_benchmark} show that the bound increases from $2.70\times10^{-4}$ at $q=0.5$ to $0.993$ at $q=3$. The theorem is a channel-structure result: it excludes exact complex-amplitude matching by lossless one-port devices whenever the ideal transmission is nonzero. It does not exclude approximate matching by lossy or multiport designs, whose energy balance and channel map must be specified separately.

Panel~\ref{fig:main-results}(f) reports the constitutive error, rather than the scattering error, of the passive dispersive model relative to the smooth finite-shell target. For the normalized parameters $\rho_{\min}=0.12a$, $\rho_1=0.22a$, $\rho_2=2.50a$, $\rho_{\max}=3.00a$, $\omega_-=0.50c/a$, $\omega_c=1.00c/a$, $\omega_+=3.00c/a$, $\Omega_0=4.00c/a$, and $\gamma=0.25c/a$, the target maximum is $N_{\max}\simeq7.15$ and Eq.~\eqref{eq:6:23} gives $\delta_{\mathrm{mat}}\simeq0.982$. This value is large; the parameter set is therefore a transparent benchmark of the calibration and error calculation, not a claim of high-fidelity broadband material emulation. The real part is calibrated to equal the target at $\omega_c$, while damping leaves a nonzero imaginary part and hence a nonzero complex error. The exact expression allows other shell, resonance, damping, and band choices to be assessed without conflating material fit with wave-level performance.

The constitutive error $\delta_{\mathrm{mat}}$ is not a scattering error. The regularized profile returns to vacuum in its core and exterior and therefore has no second asymptotic end. Its scattering matrix depends on the complete finite geometry, material dispersion and loss, interfaces, and the channels retained in the model. A device-level comparison must specify a common channel normalization and compute the full-wave scattering matrices of the ideal and regularized systems with controlled spatial, frequency, and domain convergence. Only after this channel identification is a norm such as $\delta_S(\omega)$ in Eq.~\eqref{eq:6:30} defined. Neither the constitutive fit nor the ideal-geometry scattering data substitutes for that calculation.

The principal conclusions are: (i) a global, orientation-preserving isotropic map of the full Ellis radial line; (ii) the exact impedance-matched continuum medium and its vector-Maxwell operator; (iii) the electromagnetic radial scattering problem, its spectral properties, converged multipole data, the low-frequency threshold law with coefficients validated for $L=1,2,3,4$, and the transmission and total cross-section behavior; (iv) a strictly positive reflection-mismatch bound for regular, lossless finite one-port realizations; and (v) a smooth finite-shell target together with an exact finite-band constitutive-error formula for a calibrated passive Lorentz response. The central device-level conclusion is precise: whenever the ideal two-ended geometry transmits nonzero flux, no regular, lossless finite one-port realization can reproduce its complex reflection amplitude exactly. This obstruction follows from channel structure and does not depend on the detailed finite profile. It does not extend unchanged to lossy or multiport systems, for which performance must be assessed using full-wave scattering matrices with explicitly matched channels and normalization. The Lorentz model is therefore a constitutive benchmark, not a claim of scattering-equivalent realization.

\section*{Data availability}
The scattering data are reported in Table~\ref{tab:scattering_benchmark}. All data underlying the figures and tables are openly available as ancillary files accompanying the arXiv version of this article.

\section*{Code availability}
The Python and MATLAB codes used to generate all numerical results and figures are openly available as ancillary files accompanying the arXiv version of this article, together with instructions for reproducing each figure and table.

\section*{Author contributions}
All authors contributed to the conception and development of this work, the interpretation of the results, and the writing of the manuscript. All authors read and approved the final version.

\section*{Funding}
C.F. acknowledges support from the Conselho Nacional de Desenvolvimento Cient\'ifico e Tecnol\'ogico (CNPq), Brazil (Grant No.~311781/2021-7).

\section*{Declaration of competing interests}
The authors declare no competing interests that could have influenced the work reported in this paper.

\end{document}